# TITLE

Magnetic asymmetry induced anomalous spin-orbit torque in IrMn

# AUTHOR

Jing Zhou,[1,†] Xinyu Shu,[1,†] Yaohua Liu,[2] Xiao Wang,[3] Weinan Lin,[1] Shaohai Chen,[1] Liang Liu,[1] Qidong Xie,[1] Tao Hong,[2] Ping Yang,[4,1] Bingbai Yan,[5] Xiufeng Han,[3] and Jingsheng Chen[1,6*]

1. Materials Science & Engineering, National University of Singapore, 9 Engineering Drive 1, Singapore 117575.

2. Neutron Scattering Division, Oak Ridge National Laboratory, Oak ridge, United States.

3. Institute of Physics and Beijing National Laboratory for Condensed Matter Physics, Chinese Academy of Sciences, Beijing, China.

4. Singapore Synchrotron Light Source (SSLS), National University of Singapore, 5 Research Link, Singapore 117603.

5. Department of condensed matter physics, Weizmann institute of science, Israel.

6. Suzhou Research Institute, National University of Singapore, Suzhou, China 215123.

† These authors contribute equally.

* Corresponding Author. Email: msecj@nus.edu.sg

# ABSTRACT

We demonstrate an anomalous spin-orbit torque induced by the broken magnetic symmetry in the antiferromagnet IrMn. We study the magnetic structure of three phases of IrMn thin films using neutron diffraction technique. The magnetic mirror symmetry $\mathcal{M}'$ is broken laterally in both $L1_0$-IrMn and $L1_2$-IrMn$_3$ but not γ-IrMn$_3$. We observe an out-of-plane damping-like spin-orbit torque in both $L1_0$-IrMn/permalloy and $L1_2$-IrMn$_3$/permalloy bilayers but not in γ-IrMn$_3$/permalloy. This is consistent with both the symmetry analysis on the effects of a broken $\mathcal{M}'$ on spin-orbit torque and the theoretical predictions of the spin Hall effect and the Rashba-Edelstein effect. In addition, the measured spin-orbit torque efficiencies are 0.61±0.01, 1.01±0.03 and 0.80±0.01 for the $L1_0$, $L1_2$ and γ phases, respectively. Our work highlights the critical roles of the magnetic asymmetry in spin-orbit torque generation.

# I. INTRODUCTION

The current induced spin-orbit torque (SOT) has been intensively researched since it can switch magnetization electrically, which is crucial for developing the next-generation magnetic memories [1-5]. After almost a decade since its conception, sizable SOT can be sourced from a large variety of materials, such as the topological insulator [6], heavy metal (HM) [4,7], antiferromagnets (AFM) [8-10], ferromagnetic (FM) semiconductor [11] and FM trilayer [12]. The spin Hall effect (SHE) [4] and the Rashba-Edelstein effect (REE) [5] are the two widely accepted models for the SOT generation. In both scenarios, SOT arises from the charge-to-spin conversion, where a transverse spin polarization ($s_y \| y$) is induced by a longitudinal charge current ($J_c \| x$). This spin polarization exerts an in-plane damping-like SOT ($\boldsymbol{\tau_{DL}}$) on an FM with in-plane magnetization such as the $Ni_{81}Fe_{19}$ [also known as permalloy (Py)] [1-3]. In emerging studies on SOT, however, an out-of-plane spin polarization ($s_z \| z$) is explored since it is favorable for switching a perpendicularly magnetized FM [12-16], which is more suitable for high-density applications. In this regard, the 2D material $WTe_2$ has attracted lots of attention since an out-of-plane $\tau_{DL}$, which is equivalent to the effects of an out-of-plane spin polarization, is observed in a $WTe_2$/Py bilayer [13]. This anomalous torque is attributed to the broken interfacial crystal mirror symmetry ($\mathcal{M}$) of $WTe_2$. Ever since then, the crystal-asymmetry-controlled SOT has been found in other hexagonal crystal systems, such as $NbSe_2$ [17] and $MoTe_2$ [18,19]. Magnetic moments, when ordered, can lower the symmetry of a material [20, 21]. Thus, a natural question is whether the symmetry breaking introduced by the magnetic moments can induce a similar torque.

In our previous report we found the staggered magnetic moments in $L1_0$-IrMn (001) films align with the [111] direction [22], as illustrated in Fig. 1(a). We notice that in this magnetic structure there is only one magnetic mirror symmetry ($\mathcal{M}'$) parallel to the (-110) plane and no two-fold rotational invariance, as illustrated in Fig. (1b). Since $\mathcal{M}'$ is not found in the (110) plane, the lateral magnetic symmetry is said to be broken. Moreover, $L1_0$-IrMn has a high crystal symmetry (*P4/mmm*) and its (001) plane has a four-fold rotational symmetry. This makes $L1_0$-IrMn a good choice for investigating the magnetic asymmetry induced SOT because its crystal symmetry is intact while its magnetic symmetry is broken. A previous study suggests that the SOT induced by broken mirror symmetry would exhibit a strong dependence on the direction of the electric field $E$ [13]. However, it would not be so straightforward to observe such dependence experimentally in $L1_0$-IrMn due to the existence of AFM domains. Thin films of AFM, in general, are likely to be multi-domain in their ground states [23-26]. In our previous study [22], even we managed to obtain a high-quality epitaxial film, L1$_0$-IrMn still appeared as a twin-domain structure, which would have eliminated the differences between $E \parallel \mathcal{M}'$ and $E \perp \mathcal{M}'$. Therefore, the SOT induced by the broken $\mathcal{M}'$ has to be evaluated otherwise.

In this work, we exploit the IrMn films with different phases, which allow us to separate the magnetic asymmetry (broken $\mathcal{M}'$) from specific magnetic structure. We show that a phase change from L1$_0$-IrMn to L1$_2$-IrMn$_3$ leads to a different magnetic structure but with the same lateral magnetic asymmetry. Using the spin-torque ferromagnetic resonance (ST-FMR) technique, we measure the SOT efficiencies of IrMn thin films, which are 0.61±0.01, 1.01±0.03 and 0.80±0.01 for the L1$_0$, L1$_2$ and γ phases, respectively. We demonstrate an anomalous SOT (out-of-plane damping-like) in both L1$_0$-IrMn/Py and L1$_2$-IrMn$_3$/Py bilayers. This is compatible with the symmetry analysis about the effects of a broken $\mathcal{M}'$ on SOT. We also show that the microscopic origin of the observed anomalous SOT arises from an out-of-plane spin polarization, which is consistent with theoretical predictions from both the SHE and REE perspectives. In contrast, the anomalous SOT is not observed in the γ-IrMn$_3$/Py bilayer, which we attribute to the lack of magnetic asymmetry due to disordered magnetic structure in γ-IrMn$_3$. This paper is organized as the following: Sec. II shows the methods used in this work. In Sec. III, we present the major experimental results including the crystal (Sec. IIIA) and magnetic (Sec. IIIB) structures of IrMn and the observed SOT (Sec. IIIC). In Sec. IV, we discuss the origins of the anomalous SOT, considering the intrinsic effects of $\mathcal{M}'$ on SOT (Sec. IVA-B) and its microscopic origin (Sec. IVC), followed by a conclusion in Sec. V.

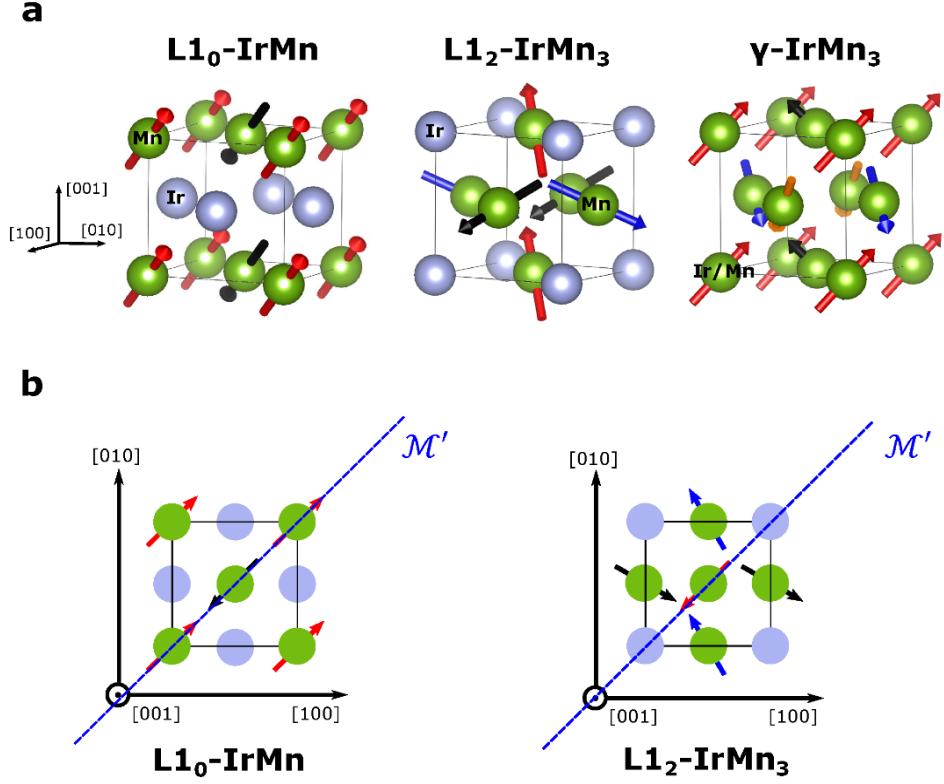

FIG. 1. (a) Unit cells of L1$_0$-IrMn, L1$_2$-IrMn$_3$ and γ-IrMn$_3$. (b) The magnetic mirror symmetry ($\mathcal{M}'$) in the (001) plane of the L1$_0$-IrMn and L1$_2$-IrMn$_3$. Arrows indicate the magnetic moments with their orientations differentiated by colors.

## II. METHODS

### A. Material fabrication

Three types of samples were fabricated in this work, namely the IrMn(22)/Py(13) bilayer with IrMn in three different phases, the L1$_0$-IrMn(22)/Cu($t_{Cu}$)/Py(13) trilayer and the Pt(10)/Py(10) bilayer. Numbers in the parentheses are thicknesses in nanometer. The IrMn (001) layers were deposited epitaxially on KTaO$_3$ (001) substrate using the DC magnetron sputtering technique at a base pressure of less than 2×10$^{-8}$ Torr. They were co-sputtered using a Ir$_{40}$Mn$_{60}$ target and a Mn target, where the atomic concentration of Ir and Mn were adjusted by the sputtering powers of the two targets. The deposition temperatures for the L1$_0$, L1$_2$ and γ phases were 720 ˚C, 640 ˚C and 320 ˚C, respectively. The measured atomic concentration of Mn for the L1$_0$, L1$_2$ and γ phases are 46.3%, 67.5% and 70.9%, respectively. After the samples cooled down to room temperature, the Cu and Py layers were deposited. The sample of Pt/Py was deposited on thermally oxidized silicon substrate at room temperature. All samples were protected by a 2-nm SiO$_2$ layer. For ST-FMR measurement, the above bilayers and trilayer were patterned into microstrips of 30 μm ×50 μm using a combination of photolithography and ion-beam etching. An electrode of Ti (5)/Cu (100) was deposited using a thermal evaporator.

## B. ST-FMR measurement

The SOT generated by IrMn was examined by the established ST-FMR technique [4,8,22]. Fig. 2(a) shows the ST-FMR experimental set-up schematically. A microwave (or equivalently the current $J_c$) sourced by a signal generator was applied along the microstrip. Then a rectifying voltage ($V_{mix}$) was produced when an in-plane external magnetic field ($H$) was swept at an angle ($\phi_H$) with respect to $J_c$. The ST-FMR measurement was modulated using a sine function of low frequency. The modulated $V_{mix}$ was collected using a lock-in amplifier and fitted using [4,8,22]

$$V_{mix} = V_S \frac{\Delta H^2}{\Delta H^2+(H-H_{res})^2} + V_A \frac{\Delta H(H-H_{res})}{\Delta H^2+(H-H_{res})^2}. \quad (1)$$

Here, $\Delta H$ is the linewidth and $H_{res}$ is the resonant field. In Eq. (1), the first term describe a symmetric component ($V_{sym}$) with an amplitude $V_S$, which is associated with in-plane torques ($\boldsymbol{\tau}_\parallel$); the second term describes an antisymmetric component ($V_{asy}$) with an amplitude $V_A$, which is due to out-of-plane torques ($\boldsymbol{\tau}_\perp$). Fig. 2(b) shows the typical $V_{mix}$ of L1$_0$-IrMn/Py, L1$_2$-IrMn$_3$/Py and γ-IrMn$_3$/Py bilayers measured at 9 GHz with $\phi_H$ = -35°. The data points can be well fitted by Eq. (1).

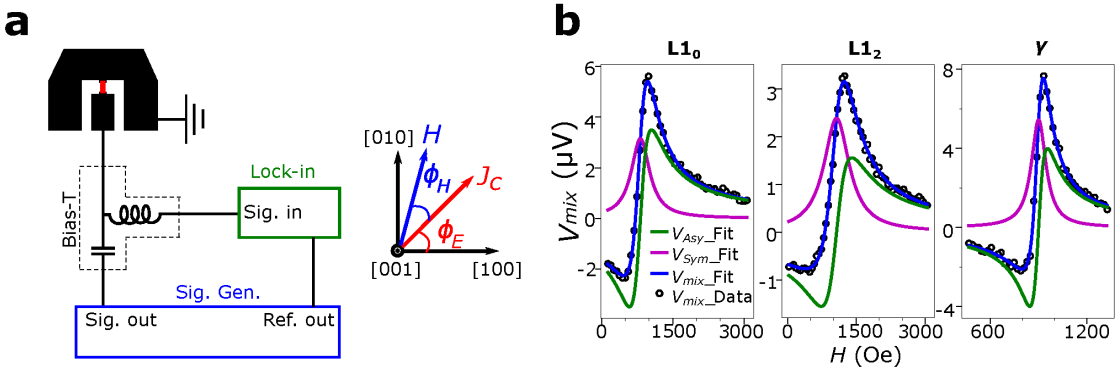

FIG. 2. (a) Experimental set-up of ST-FMR measurement. The red bar indicates the micro strip of IrMn/Py bilayer. On the right shows the coordinates system, which is defined with respect to the crystal lattice of IrMn. (b) The measured $V_{mix}$ for three phases of IrMn and their fittings at 9 GHz as a function of the external magnetic field ($H$).

The SOT efficiency, $\theta_{\parallel,m}$, which represents the magnitude of measured $\boldsymbol{\tau}_\parallel$ relative to $\boldsymbol{\tau}_\perp$, can be extracted using [4,8,22]

$$\theta_{\parallel,m} = \frac{V_S}{V_A} \frac{e\mu_0 M_s d_{IrMn} d_{Py}}{\hbar} \sqrt{1 + \frac{4\pi M_{eff}}{H_{res}}}, \quad (2)$$

where $e$, $\mu_0$, $\hbar$ are the electron charge, permeability of free space and reduced Planck constant, respectively; the thickness of the two layers are $d_{IrMn}$ and $d_{Py}$; $M_s$ and $M_{eff}$ are the saturation and effective magnetizations of the Py layer, respectively. $M_s$ is extracted from a *M-H* loop measured using a vibrating sampling magnetometer. $M_{eff}$ is extracted by fitting the in-plane Kittel equation $f = \frac{\gamma}{2\pi}\sqrt{(H_{res} + H_k)(H_{res} + H_k + 4\pi M_{eff})}$. Here, $f$ is the microwave frequency, $H_k$ is the effective in-plane magnetic anisotropic field.

### C. Principles of angle dependent ST-FMR measurement

In order to determine the symmetry of SOT, we examine the dependence of $V_S$ and $V_A$ on $\phi_H$. This approach is based on the Landau-Lifshitz-Gilbert-Slonczewski formalism [4,27] that SOT must be one of the two types: a damping-like torque $\boldsymbol{\tau_{DL}} = \tau_{DL0}(\boldsymbol{m} \times \boldsymbol{s} \times \boldsymbol{m})$ and a field-like torque $\boldsymbol{\tau_{FL}} = \tau_{FL0}(\boldsymbol{m} \times \boldsymbol{h})$. In the polar coordinate, $\boldsymbol{m} = m_0[\cos(\theta_H)\cos(\phi_H), \cos(\theta_H)\sin(\phi_H), \sin(\theta_H)]^T$ is the magnetic moment, which is aligned to the external magnetic field direction with a small precession angle. The spin polarization is $\boldsymbol{s} = s[\cos(\theta_s)\cos(\phi_s), \cos(\theta_s)\sin(\phi_s), \sin(\theta_s)]^T$ and the current induced dynamic magnetic field is $\boldsymbol{h} = h[\cos(\theta_h)\cos(\phi_h), \cos(\theta_h)\sin(\phi_h), \sin(\theta_h)]^T$. All the in-plane angles ($\phi_h$, $\phi_H$ and $\phi_s$) are relative to the $J_c$ (defined as *x*), and the out-of-plane angles ($\theta_h$, $\theta_H$ and $\theta_s$) are relative to the film plane (*x-y* plane). The $\boldsymbol{\tau_{DL}}$ and the $\boldsymbol{\tau_{FL}}$ are differentiated by their definitions, i.e. $\boldsymbol{\tau_{DL}}$ ($\boldsymbol{\tau_{FL}}$) is even (odd) in $\boldsymbol{m}$. However, the $\boldsymbol{\tau_\parallel}$ differs from the $\boldsymbol{\tau_\perp}$ by their orientations relative to the film normal (defined as *z*).

In the bilayers dominated by the SHE, such as the Pt/Py, both the current-induced Oersted field and the spin polarization are along the *y* direction when $J_c$ is along the *x* direction [4]. $\boldsymbol{m}$ usually has two in-plane components in ST-FMR measurement. Thus, $\boldsymbol{h} = h(0,1,0)^T$, $\boldsymbol{m} = m_0[\cos(\phi_H), \sin(\phi_H), 0]^T$, and $\boldsymbol{s} = s(0,1,0)^T$. Then we have

$$\boldsymbol{\tau_{FL}} = \tau_{FL0}(\boldsymbol{m} \times \boldsymbol{h}) = \tau_{FL0} m_0 h \begin{pmatrix} 0 \\ 0 \\ \cos(\phi_H) \end{pmatrix}, \quad (3)$$

$$\boldsymbol{\tau_{DL}} = \tau_{DL0}(\boldsymbol{m} \times \boldsymbol{s} \times \boldsymbol{m}) = \tau_{DL0} s m_0^2 \begin{pmatrix} -\sin(\phi_H)\cos(\phi_H) \\ \cos^2(\phi_H) \\ 0 \end{pmatrix}. \quad (4)$$

However, in order to examine the direction of the SOT, the coordinate system must be redefined with respect to $m$, i.e. to multiply a rotation matrix

$$R_m = \begin{pmatrix} \cos(\phi_H) & \sin(\phi_H) & 0 \\ -\sin(\phi_H) & \cos(\phi_H) & 0 \\ 0 & 0 & 1 \end{pmatrix}. \quad (5)$$

As a result, $\tau_{FL}$ remains the same but

$$\begin{pmatrix} \cos(\phi_H) & \sin(\phi_H) & 0 \\ -\sin(\phi_H) & \cos(\phi_H) & 0 \\ 0 & 0 & 1 \end{pmatrix} \tau_{DL} = \tau_{DL0} s m_0^2 \begin{pmatrix} 0 \\ \cos(\phi_H) \\ 0 \end{pmatrix}. \quad (6)$$

Here, $\tau_\parallel$ only has the $\tau_{DL}$ component and $\tau_\perp$ only has the $\tau_{FL}$ component, both depend only on $\cos(\phi_H)$. Therefore, both $V_S$ and $V_A$ depend on $\sin(2\phi_H)\cos(\phi_H)$, where the additional $\sin(2\phi_H)$ is due to the anisotropic magnetoresistance (AMR) effect [4,11,13]. In the unusual event of an out-of-plane spin polarization, we have $h = h(0,1,0)^T$, $m = m_0[\cos(\phi_H), \sin(\phi_H), 0]^T$ but $s = s[0, \cos(\theta_s), \sin(\theta_s)]^T$. Following the same procedure in Eq. (3) to (6), we have

$$\tau_{DL} = \tau_{DL0} s m_0^2 \begin{pmatrix} 0 \\ \cos(\phi_H)\cos(\theta_s) \\ \sin(\theta_s) \end{pmatrix}, \quad (7)$$

$$\tau_{FL} = \tau_{FL0} m_0 h \begin{pmatrix} 0 \\ 0 \\ \cos(\phi_H) \end{pmatrix}. \quad (8)$$

In this case, $\tau_{DL}$ has both $\tau_\parallel$ and $\tau_\perp$ components. Alternatively speaking, $\tau_\parallel$ still depends only on $\cos(\phi_H)$ but $\tau_\perp$ depends on $[\cos(\phi_H) + \sin(\theta_s)]$, which after multiplying the $\sin(2\phi_H)$ term due to AMR effect, create a new angular dependence on $\phi_H$ of $V_A$.

TABLE I. Dependence of $V_S$ and $V_A$ on $\phi_H$. IP stands for in-plane, OP stands for out-of-plane. The subscripts on $s$, $h$ and $m$ indicate the present components, e.g. $h_{yz}$ indicates the current-induced field has both $y$ and $z$ components. Other coefficients before the *sine* and *cosine* terms are omitted to highlight the angular dependence.

| Case | Configuration | Dependence on $\phi_H$ | | Origin[a] |
|---|---|---|---|---|
| | | $V_A (\propto \tau_\perp)$ | $V_S (\propto \tau_\parallel)$ | |
| 1 | $s_y, h_y, m_{xy}$ | $\cos(\phi_H)$ | $\cos(\phi_H)$ | Classic SHE |
| 2 | $s_y, h_{yz}, m_{xy}$ | $\cos(\phi_H)\cos(\theta_h)$ | $\cos(\phi_H) - \sin(\theta_h)$ | OP field |
| 3 | $s_y, h_{xy}, m_{xy}$ | $\sin(\phi_h - \phi_H)$ | $\cos(\phi_H)$ | IP field |
| 4 | $s_{xy}, h_y, m_{xy}$ | $\cos(\phi_H)$ | $\sin(\phi_H - \phi_s)$ | IP spin polarization |
| 5 | $s_{yz}, h_y, m_{xy}$ | $\cos(\phi_H) + \sin(\theta_s)$[b] | $\cos(\phi_H)\cos(\theta_s)$ | OP spin polarization |
| 6 | $s_y$, $h_y$, $m_{xyz}$ | $\sin(\phi_H)\sin(\theta_m)$ $+ \cos(\phi_H)$ | $\sin(\phi_H)\sin(\theta_m)$ $+ \cos(\phi_H)$ | OP magnetic moment |

[a]in additional to the configuration of classic SHE.

[b]this is the only $\phi_H$ independent term for $V_A$.

The results of Eq. (3), (6) to (8) and other possible factors influencing the angular dependence of $V_S$ and $V_A$ are summarized in Table I. Though these contributions might not be exhaustive, we briefly discuss their possible origins. Case 2 happens if the electric contact between the G-S-G RF probe and the electrode is not homogeneous, leading to a net $h_z$ [28]. A possible cause of case 3 is a Dresselhaus-like field ($h_x$). Case 4 describes an unusual spin polarization that is collinear with $J_c$. Case 6 can arise from an interfacial exchange coupling between an FM and an AFM [29]. The rotation matrix $R_m$ is different for case 6 due to the presence of $m_z$. In this case, the classification of $\boldsymbol{\tau_{DL}}$ and $\boldsymbol{\tau_{FL}}$ into $\boldsymbol{\tau_\parallel}$ and $\boldsymbol{\tau_\perp}$ is trivial since $\boldsymbol{m}$ is between the in-plane and out-of-plane directions.

Considering all cases in Table I and omitting the $\sin(2\phi_H)$ contribution from AMR, both $V_S$ and $V_A$ consist of a $\cos(\phi_H)$ term, a $\sin(\phi_H)$ term and a term independent of $\phi_H$, leading to their general expression

$$V_S = \sin(2\phi_H - 2\phi_0)[S_A \cos(\phi_H - \phi_0) + S_B \sin(\phi_H - \phi_0) + S_C], \quad (9)$$

$$V_A = \sin(2\phi_H - 2\phi_0)[A_A \cos(\phi_H - \phi_0) + A_B \sin(\phi_H - \phi_0) + A_C], \quad (10)$$

where $\phi_0$ is a phase correction to $\phi_H$. Therefore, based on Table I, Eq. (9) and (10), the various contributions to SOT can be differentiated by performing the ST-FMR measurement at different magnetic field directions.

## III. EXPERIMENTAL RESULTS

### A．Crystal structures of IrMn

Since IrMn exists in different crystal phases with different magnetic structures, it is of prime importance to verify the phases of IrMn before investigating its SOT generation. Fig. 3(a) shows the analytical rules used in this work to determine the phases of IrMn using x-ray diffraction (XRD) technique. In a θ-2θ scan along the (001) direction, the presence of the 001 peak indicates the presence of chemically ordered $L1_0$-IrMn or $L1_2$-IrMn$_3$, whereas only the 002 peak is expected in the chemically disordered γ-IrMn$_3$. The $L1_2$-IrMn$_3$ is differentiated from the $L1_0$-IrMn by the presence of (103) spot on the reciprocal space mapping (RSM). For polycrystalline IrMn (p-IrMn) deposited at room temperature, peaks can be hardly detected from the above XRD tests due to its small grain size and poor crystallinity, and it only produces ring-like features in RSM when the film is sufficiently thick. Fig. 3(b) shows the typical results of the θ-2θ scans. The $L1_0$ and $L1_2$ phases have all peaks in the {001} family and no other peaks, indicating high crystallinity and good (001) texture. In addition, both the $L1_2$ and γ phases are slightly strained due to lattice mismatch with the substrate, since $[a(KTaO_3) = 3.989$ Å$] > [a(L1_2,\gamma\text{-}IrMn_3) = 3.785$ Å$]$ [30]. This is seen from the rightward shift of the 002 peak from the reference bulk value in Fig. 3(b), indicating shorter lattice parameter $c$, consistent with the usual effects of a tensile strain. The typical RSMs are shown in Fig. 3(c), where the (113) spot is common for all three phases but only the $L1_2$ phase has the (103) spot.

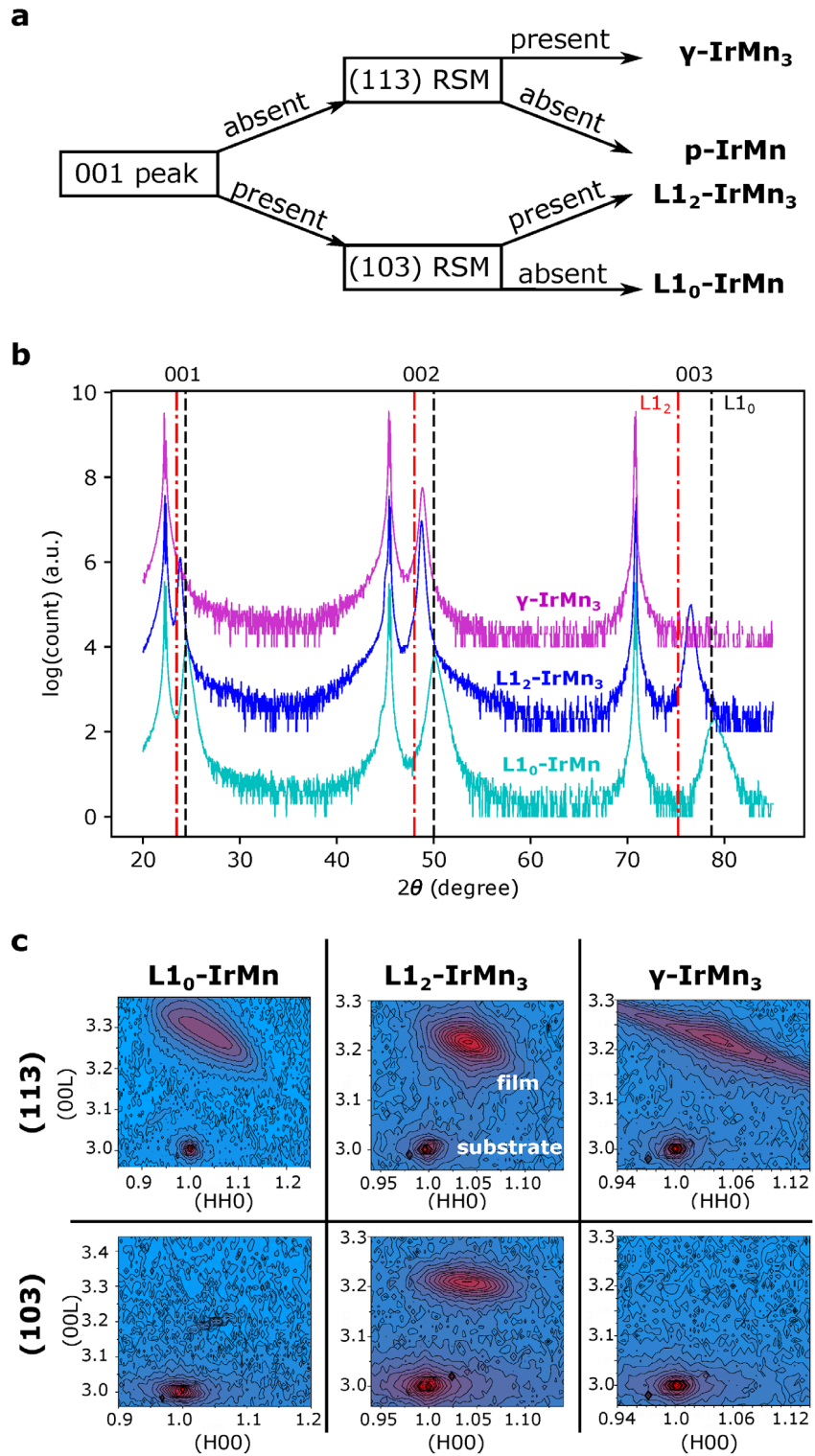

FIG. 3. (a) Guiding rules for determining the phases of IrMn by using the XRD technique. p-IrMn refers to polycrystalline IrMn. (b) XRD θ-2θ scan of three phases of IrMn along (001) direction. The dashed vertical lines show the reference peak positions [9, 30] for bulk IrMn with the Cu-K$_{\alpha 1}$ radiation equivalent. (c) Typical RSMs of IrMn in different phases near (103) and (113) spots. The RSMs are indexed based on the reciprocal lattice of KTaO$_3$.

### B. Magnetic structures of IrMn

The three phases of IrMn are further studied by examining their magnetic structure using neutron diffraction technique. The collinear magnetic structure of $L1_0$-IrMn shown in Fig. 1(a) is proposed from our previous work [22] based on the combined results from neutron diffraction and ST-FMR. The magnetic structure of $L1_2$-IrMn$_3$ is first examined by scanning a large reciprocal space using a time-of-flight quasi-Laue diffractometer. Fig. 4(a) shows the RSM of (H0L) (left) and two typical line cuts (right). The presence of a lot of diffraction spots of the film indicates good magnetic order. The ring-like features originate from the sample environment such as the contribution of the sample mount. The absence of the (002) and (202) spots on the (H0L) RSM and the absence of the (111) spot on the (HHL) RSM (not shown) indicate a good $L1_2$ order [31]. Given the good magnetic order, we do not observe any superlattice spot, for example (±0.5 0 0), (0.5 0 0.5) and (1 0 0.5). This is another indicator of the triangular spin structure of the $L1_2$ phase [31]. The magnetic structure of $L1_2$-IrMn$_3$ is also studied by measuring the integrated intensities of selected peaks by θ-2θ scans using a triple-axis spectrometer. Fig. 4(b) shows the 100 and 110 peaks and their Gaussian fits. Their integrated intensities are $I_{100} = 450 \pm 20$ and $I_{110} = 570 \pm 20$, respectively. Among the possible magnetic structures of $L1_2$-IrMn$_3$ proposed previously [32], the one shown in Fig. 1(a) is the only viable model. Other models result in either satellite peaks due to doubled unit cell length along *c*-axis or weaker integrated intensity of the 110 peak than the 100 peak (i.e. $I_{110}<I_{100}$). Both are negative in our results. Therefore, based on the above evidences, we have verified the magnetic structure of the $L1_2$ phase, which agrees with previous studies [8,30-32]. On the other hand, we have not observed any diffraction spot for the γ-IrMn$_3$, inconsistent with the magnetic structure proposed previously [30], as shown in Fig. 1(a). This indicates that our γ-IrMn$_3$ does not have a sufficiently ordered magnetic structure although the observed crystal structure positively identifies a γ phase. Based on the results of neutron diffraction, we illustrate in Fig. 1(b) that the $L1_2$-IrMn$_3$ also has just one magnetic mirror symmetry $\mathcal{M}'$ parallel to the (-110) plane, although its magnetic structure is markedly different from the $L1_0$-IrMn. Correspondingly, the lateral magnetic symmetry in $L1_2$-IrMn$_3$ is broken in the same manner as $L1_0$-IrMn due to the lack of $\mathcal{M}'$ in other directions. In γ-IrMn$_3$, $\mathcal{M}'$ is absent and therefore no magnetic asymmetry is identified.

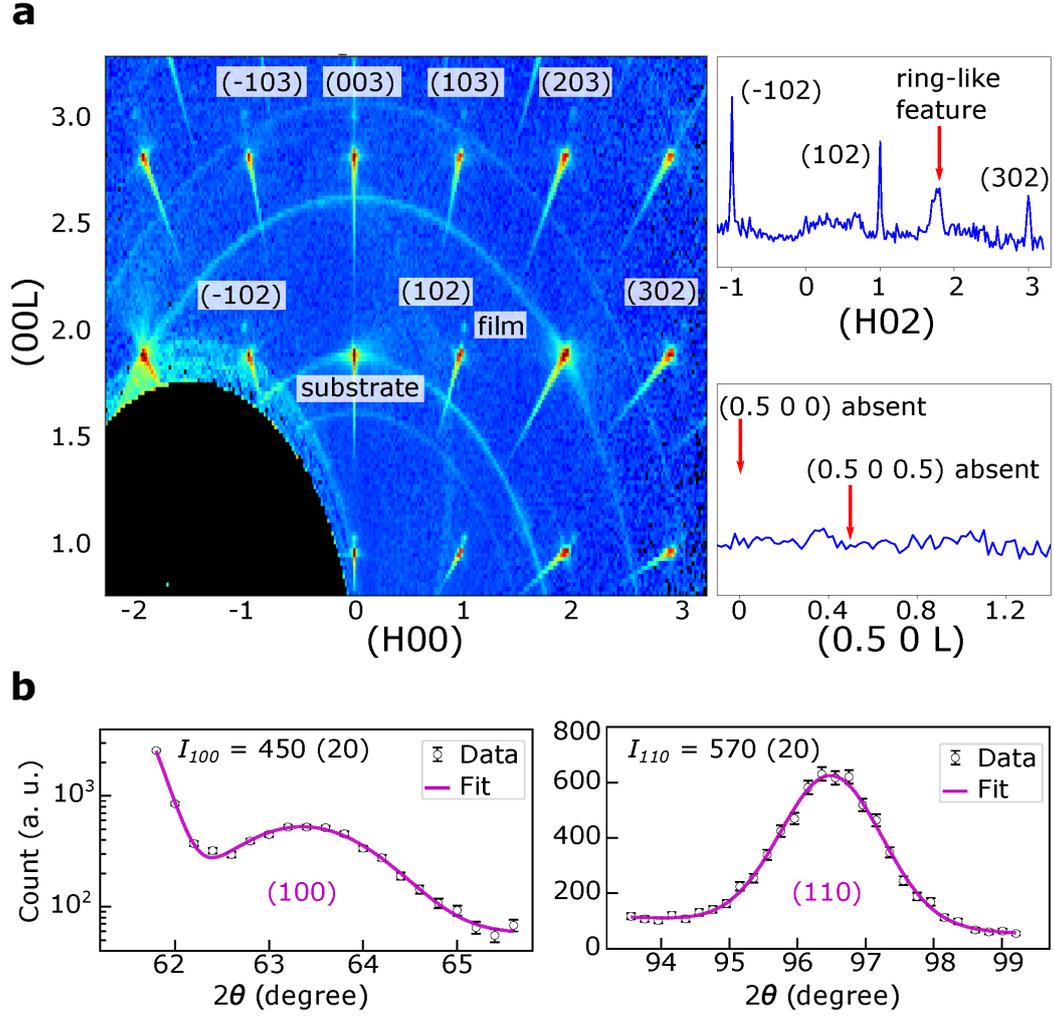

FIG. 4. (a) Left: large-area neutron diffraction RSM of (H0L) of $L1_2$-IrMn$_3$. Right: line cuts of (H02) and (0.5 0 L) of the RSM on the left. The peaks are indexed in the reciprocal lattice of $L1_2$-IrMn$_3$. (b) θ-2θ scans on $L1_2$-IrMn$_3$ along (100) and (110) directions. $I_{100}$ and $I_{110}$ are the integrated intensity adjusted by absorption corrections.

### C. Spin-orbit torque in IrMn

The SOT efficiency $\theta_{\parallel,m}$ of three phases of IrMn are summarized in Fig. 5(a), where minimal variation is observed across 8-12GHz. The $\theta_{\parallel,m}$ (averaged over 8-12GHz) are 0.61±0.01, 1.01±0.03 and 0.80±0.01 for the $L1_0$, $L1_2$ and γ phases, respectively. These values are substantially larger than those from previous reports [8, 9, 33-35], which we attribute to the different crystal and magnetic ordering. Our IrMn thin films have high crystallinity with the ordered magnetic structures in the $L1_0$ and $L1_2$ phases. As a results, the averaging effect [9,22] due to randomly oriented crystallites and magnetic domains is expected to be much smaller, leading to greater values of measured SOT efficiency. It has been reported that $\theta_{\parallel,m}$ is strongly affected by the electrical resistivity ($\rho$) [7, 36]. An effective spin Hall conductivity ($\sigma_s$) is therefore calculated using $\sigma_s = \frac{\theta_{\parallel,m}}{\rho}$ [36], which is typically $\sim 10^3 \frac{\hbar}{e} \frac{S}{cm}$ in our work. Referring to Fig. 5(b), L1$_0$-IrMn and L1$_2$-IrMn$_3$ have similar $\rho$. However, the $\theta_{\parallel,m}$ of $\gamma$-IrMn$_3$ is profoundly inflated by its high $\rho$, and its $\sigma_s$ is actually close to that of L1$_0$-IrMn. Interestingly, comparing the L1$_0$ and L1$_2$ phases, the ratio of measured $\sigma_s$ ($\frac{3.55}{6.12} = 0.58$) is very close to the ratio of calculated intrinsic $\sigma_s$ ($\frac{102}{165} = 0.62$) {$\sigma_s$ values adapted from Ref. [8, 22]}. This implies that the intrinsic SHE plays an important role in the SOT generated by IrMn.

The SOT efficiency of $\gamma$-IrMn$_3$ has not been reported before to the best of our knowledge. Unlike the magnetic-structure-enhanced SOT efficiency in L1$_0$-IrMn and L1$_2$-IrMn$_3$, the reasons behind the relatively large SOT efficiency and spin Hall conductivity in $\gamma$-IrMn$_3$ are still unclear. However, we find that the mixed experimental evidences from a previous work might support our observation. A giant SOT efficiency of 0.35 has been claimed for L1$_2$-IrMn$_3$ (001) [8], but only the 002 peak from the XRD results is observed without the superlattice 001 peak. This makes the phase of the IrMn$_3$ film under investigation ambiguous since the presence of only the 002 peak corresponds to a $\gamma$-IrMn$_3$.

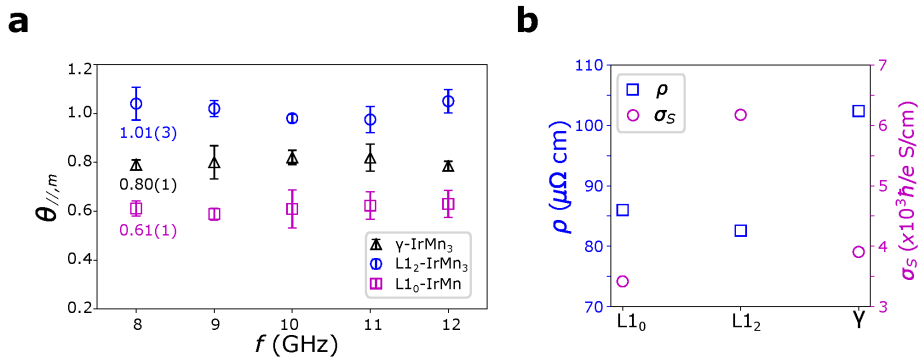

FIG. 5. (a) The measured SOT efficiencies ($\theta_{\parallel,m}$) of the three phases in the frequency range of 8-12 GHz with $\phi_H$ = -35° and $\phi_E$ = 45°. (b) The electric conductivity ($\rho$) and spin Hall conductivity ($\sigma_s$) of the three phase of IrMn.

In Sec. IIC we show that both the symmetry voltage amplitude $V_S$ and the antisymmetric voltage amplitude $V_A$ depends on $\sin(2\phi_H)\cos(\phi_H)$ in a bilayer dominated by the in-plane transverse spin polarization ($s_y$) and Oersted field, such as Pt/Py. Thus, $V_{mix}$ is expected to be centrosymmetric with $H$. However, Fig. 6(a) shows that $V_{mix}$ measured from L1$_0$-IrMn/Py is not a perfect inversion when $H$ is reversed, where $V_{sym}$ remains the same but $V_{asy}$ is markedly different, indicating the presence of an additional out-of-plane torque. Therefore, we perform the ST-FMR measurement with $\phi_H$ varying from -90° to 270° and fit $V_S$ and $V_A$ using Eq. (9) and (10). Fig. 6(b) shows the typical results for $L1_0$-IrMn/Py. The associated fitting parameters are $S_A$ = -3.9 ± 0.3 μV, $S_B$ = 0.06 ± 0.7 μV, $S_C$ = -0.08 ± 0.4 μV, $A_A$ = -8.1 ± 0.6 μV, $A_B$ = -0.2 ± 0.7 μV and $S_C$ = -0.6 ± 0.2 μV. All parameters with a larger fitting error than value are considered zero within experimental accuracy. Therefore, $V_S$ shows a usual shape with only the $S_A$ contribution. In contrast, $V_A$ has a non-zero $A_C$ term on top of the normal $A_A$ contribution. The fitted components of $V_A$ are displayed separately in Fig. 6(c). We also observe similar dependence of $V_{mix}$ on $\phi_H$ in $L1_2$-IrMn$_3$/Py bilayer and its $V_A$ also has an $A_C$ contribution, as demonstrated in Fig. 6(d). Among the possible components of $V_A$ in Table I, $\sin(\theta_S)$ under Case 5 is the only term that shares the similar angular dependence of $A_C$, which implies the presence of an out-of-plane spin polarization ($s_z$). Moreover, according to Eq. (7), the out-of-plane torque associated with $\sin(\theta_S)$ (tagged as $\tau_{AC}$) is damping-like since it is even in ***m***.

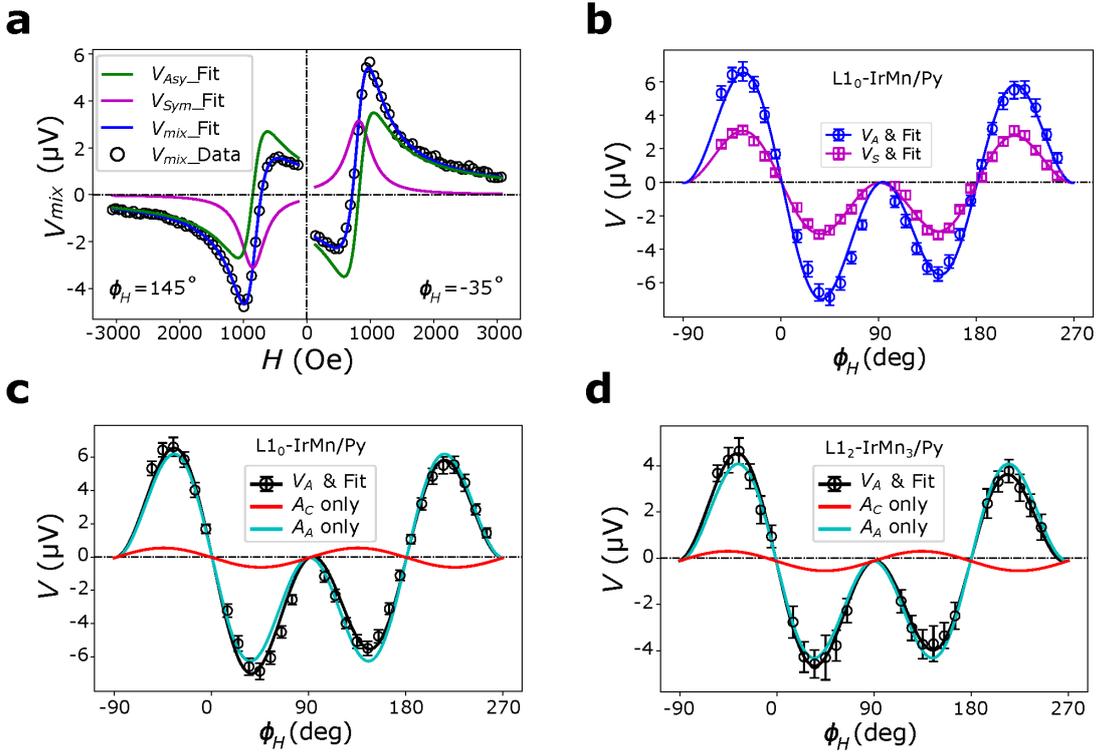

FIG. 6. (a) Raw data of measured voltage ($V_{mix}$) and its fit for $L1_0$-IrMn/Py at 9 GHz with $\phi_H$ = -35° and $\phi_H$ = 145°. (b) $V_S$, $V_A$ and their fits as a function of $\phi_H$ for $L1_0$-IrMn/Py. (c) $V_A$ and its fitted component for $L1_0$-IrMn/Py and (d) $L1_2$-IrMn$_3$/Py. $\phi_E$ = 45° in all plots.

We perform two more tests to study $\tau_{AC}$. First, we investigate the dependence of $\tau_{AC}$ on the in-plane electric field direction defined by $\phi_E$ ($\phi_E = 0°$ || [100] of the L1$_0$-IrMn lattice). This is achieved by patterning the microstrip along different in-plane directions on the same sample. Fig. 7(a) and (b) show the ST-FMR voltages and their fittings for $\phi_E = 0°$ and $\phi_E = 90°$ in L1$_0$-IrMn/Py, respectively. The results for both $\phi_E = 0°$ and $\phi_E = 90°$ are similar to those for $\phi_E = 45°$ in Fig. 6(b). The magnitudes of $A_C$ for $\phi_E = 0°$, 45° and 90° are normalized and summarized in Fig. 7(e). The ratio of $A_C/A_A$ for L1$_0$-IrMn is less than 10% and it is roughly independent of $\phi_E$. This is consistent with our prediction that the twin domains of L1$_0$-IrMn would nullify the directionality of $\tau_{AC}$. Second, we insert a thin Cu layer between L1$_0$-IrMn and Py to break their strong exchange coupling. Fig. 7(c) and (d) show that the measured $V_A$ for $t_{Cu} = 0.5$ nm and $t_{Cu} = 1$ nm have a similar shape to that for $t_{Cu} = 0$ nm in Fig. 7(a), although the magnitude of $V_A$ substantially increases after Cu is inserted due to lower impedance mismatch. The extracted values of $A_C/A_A$ are summarized in Fig. 7(f). Essentially, $A_C/A_A$ is also independent of the Cu spacer thickness in L1$_0$-IrMn (22)/Cu($t_{Cu}$)/Py (13). This indicates that the interface between L1$_0$-IrMn and Py is not likely to account for $\tau_{AC}$, which appears to originate from the AFM layer only. Notice that we have estimated the shunting effect due to the Cu spacer, which is negligible [22]. As a comparison, Fig. 7(e) and (f) also display $S_A/A_A$, which depends on both $\phi_E$ and $t_{Cu}$, implying the strong influence of exchange coupling. This further illustrates that the nature of $\tau_{AC}$ is different from the usual SOT.

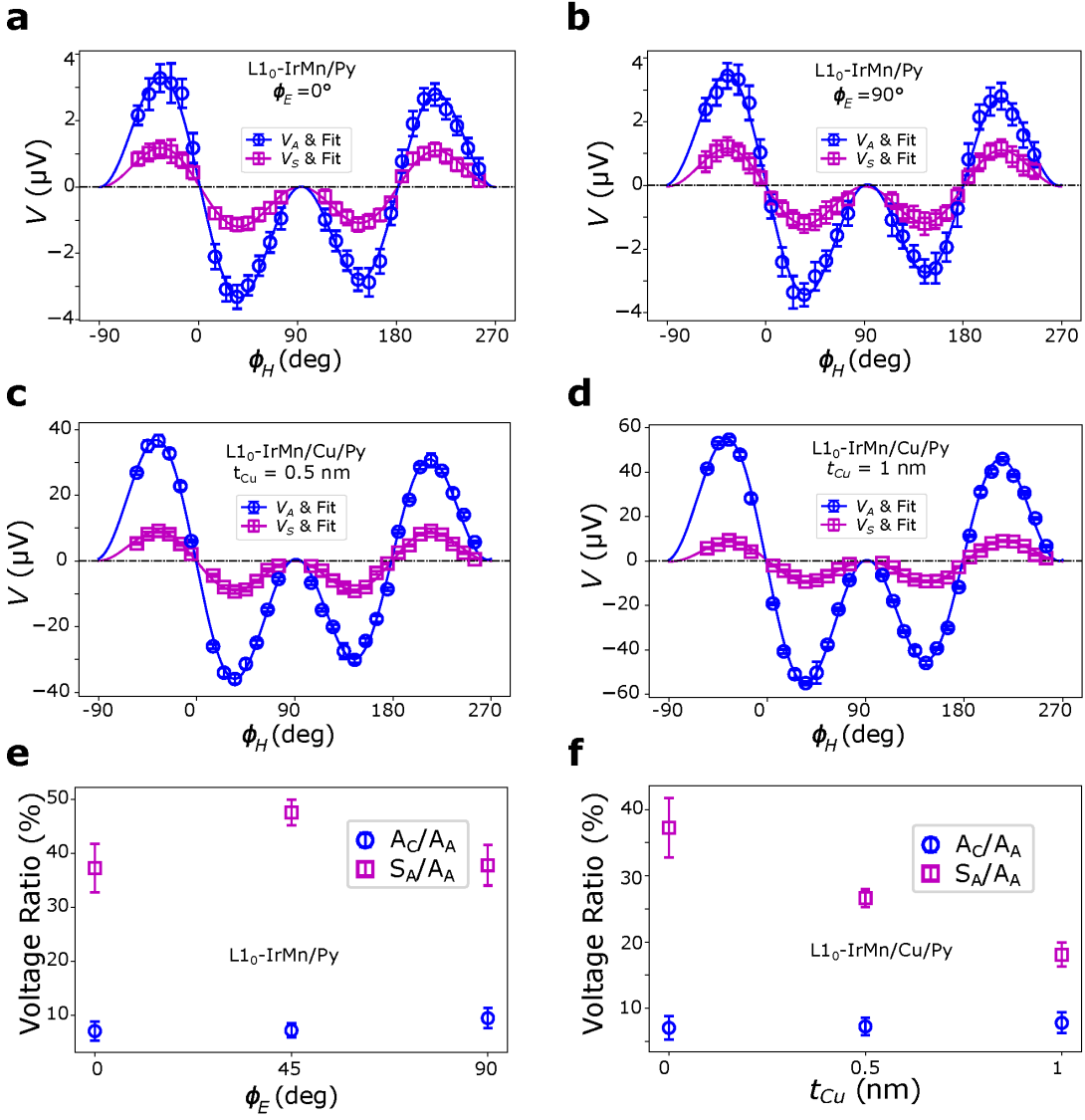

FIG. 7. (a) $V_S$, $V_A$ and their fits as a function of $\phi_H$ for L1$_0$-IrMn/Py with $\phi_E = 0°$ and (b) $\phi_E = 90°$. (c) $V_S$, $V_A$ and their fits as a function of $\phi_H$ for L1$_0$-IrMn/Cu($t_{Cu}$)/Py with $t_{Cu}$=0.5 nm and (d) $t_{Cu}$=1 nm. $\phi_E = 0°$ in (c) and (d). (e) Variation of normalized $S_A$ and $A_C$ with $\phi_E$ for L1$_0$-IrMn/Py and (f) $t_{Cu}$ for L1$_0$-IrMn/Cu($t_{Cu}$)/Py.

## IV. DISCUSSION

### A. Symmetry analysis on SOT

The common $\tau_{AC}$ in both L1$_0$-IrMn and L1$_2$-IrMn$_3$ can be interpreted from symmetry analysis about the intrinsic effects of $\mathcal{M}'$ on SOT. The magnetic mirror symmetry $\mathcal{M}'$ differs from the crystal mirror symmetry $\mathcal{M}$. $\mathcal{M}'$ contains a time reversal symmetry $T$, i.e. $\mathcal{M}' = \mathcal{M} * T$ [20]. An axial vector like the magnetic moment is odd in $T$ whereas a polar vector like the electric field is even in $T$. As a result, the behaviors of a polar vector and an axial vector are opposite (same) under $\mathcal{M}$ ($\mathcal{M}'$). This is illustrated in Fig. 8.

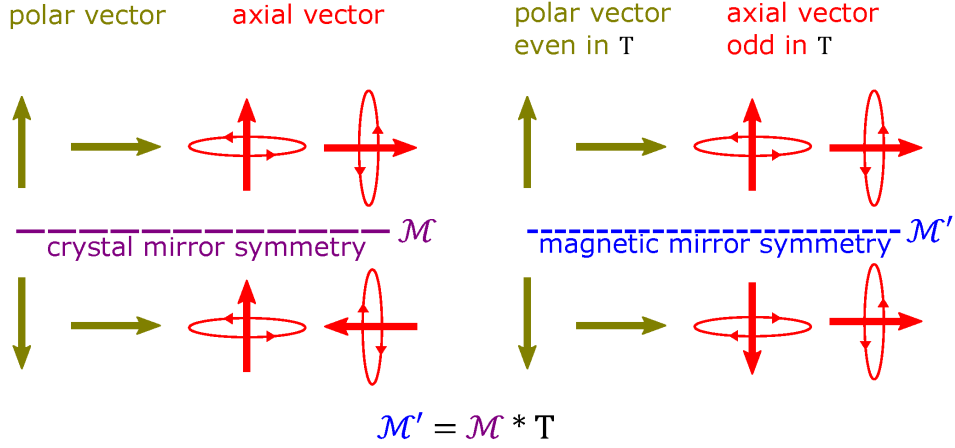

FIG. 8. Effects of symmetry operation $\mathcal{M}$ and $\mathcal{M}'$ on polar and axial vectors. $\mathcal{M}$ and $T$ represents the mirror symmetry and time reversal symmetry.

Based on these relations and inspired by the approach to analyze crystal mirror symmetry in Ref. [13], we scrutinize the effects of the magnetic mirror symmetry $\mathcal{M}'$ on three parameters: (1) the in-plane and out-of-plane SOTs, which are collectively denoted as $\tau$, (2) the electric field $E$ and (3) the in-plane angle $\phi_H$, which defines the direction of magnetization. Notice that $\phi_H$ is always defined relative to the direction of $E$, as specified in the Fig. 2(a). As a result, the coordinate system (indicted by $x$ in Fig. 9) relative to $\mathcal{M}$ and $\mathcal{M}'$ is changing with the direction of $E$. The effects of $\mathcal{M}'$ and $\mathcal{M}$ are compared as the following:

(1) It has been shown that $\tau$ is a pseudoscalar [13]. Therefore, it remains the same when $\mathcal{M}'$ is applied but it changes sign in the case of $\mathcal{M}$.

(2) In the cases of both $\mathcal{M}'$ and $\mathcal{M}$, $E$ is flipped if $E \perp \mathcal{M}'(\mathcal{M})$ and $E$ is unchanged if $E \parallel \mathcal{M}'(\mathcal{M})$.

(3) When $E \perp \mathcal{M}'$, $x \perp \mathcal{M}'$. As a result, $\phi_H \rightarrow \pi - \phi_H$ when $\mathbf{m}$ is reflected by $\mathcal{M}'$, as shown in Fig. 9(a). When $E \parallel \mathcal{M}'$, $x \parallel \mathcal{M}'$. As a result, $\phi_H \rightarrow -\phi_H$ when $\mathbf{m}$ is reflected by $\mathcal{M}'$, as shown in Fig. 9(b). Since $\mathbf{m}$ is an axial vector, it has to be flipped again after reflection if $\mathcal{M}$ is applied in the place of $\mathcal{M}'$. This is because the crystal mirror symmetry $\mathcal{M}$ does not contain the time reversal symmetry. Therefore, as shown in Fig. 9(c) and (d), $\phi_H \rightarrow -\phi_H$ for $E \perp \mathcal{M}$ and $\phi_H \rightarrow \pi - \phi_H$ for $E \parallel \mathcal{M}$.

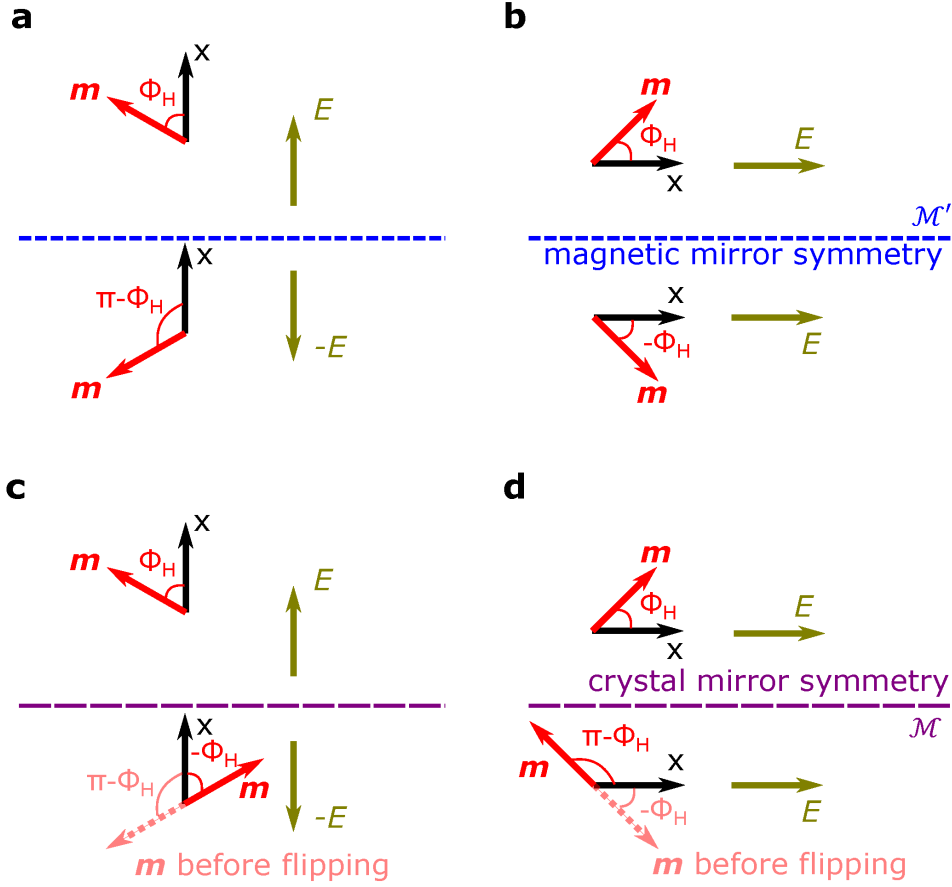

FIG. 9. (a) Effects of $\mathcal{M}'$ on $E$ and $\phi_H$ for $E \perp \mathcal{M}'$ and (b) $E \parallel \mathcal{M}'$. (c) Effects of $\mathcal{M}$ on $E$ and $\phi_H$ for $E \perp \mathcal{M}$ and (d) $E \parallel \mathcal{M}$. $x$ is always parallel to $E$, it shows the coordinate system with which $\phi_H$ is defined. The dotted arrow of light red color in (c) and (d) indicates the direction of magnetic moment $m$ after reflection by $\mathcal{M}$ but before flipping.

Since SOT is a function of electric field and magnetization direction, i.e. $\tau = \tau(E, \phi_H)$, the effects of $\mathcal{M}'$ on the three parameters pose constraints on $\tau$ if $\tau$ is Fourier expanded. The unconstrained SOT is $\tau(E, \phi_H) = E[F_0 + F_1 \cos(\phi_H) + F_2 \sin(\phi_H) + F_3 \cos(2\phi_H) + F_4 \sin(2\phi_H) + \cdots]$ [13]. For $E \perp \mathcal{M}'$ in Fig. 9(a), the net results of $E \to -E$, $\tau \to \tau$ and $\phi_H \to \pi - \phi_H$ is $\tau(-E, \pi - \phi_H) = \tau(E, \phi_H)$. Since $\tau(E) = -\tau(-E)$ for current induced SOT, we have $\tau(\phi_H) = -\tau(\pi - \phi_H)$, which reduces the unconstrained form of $\tau$ to $\tau(E, \phi_H) = E[F_1 \cos(\phi_H) + F_4 \sin(2\phi_H) + F_5 \cos(3\phi_H) + \cdots]$, where $F_0$ is not allowed. Similarly, it can be shown that for $E \parallel \mathcal{M}'$ in Fig. 9(b), $\tau(\phi_H) = \tau(-\phi_H)$, which means $\tau(E, \phi_H) = E[F_0 + F_1 \cos(\phi_H) + F_3 \cos(2\phi_H) + F_5 \cos(3\phi_H) + \cdots]$, where $F_0$ is allowed. $F_0$ carries the similar physical significance as $A_C$ and $S_C$ in the main text since it is independent of $\phi_H$. Therefore, $F_0$ rationalizes the observed out-of-plane damping-like torque. These results are summarized in

Table II and compared with the effects of $\mathcal{M}$ in WTe$_2$ [13]. Essentially, $\mathcal{M}$ and $\mathcal{M}'$ are similar that both allow the observed $A_C$ and thus $\tau_{AC}$, but they differ in the direction of $E$ with which this torque is allowed. Note that though $F_0$ is also allowed for the in-plane SOT, the measured $S_C$ is negligible. We could not identify any material-dependent microscopic origins other than the inhomogeneous contact of probe (Sec. IIC), which we have minimized with the best of our engineering effort. The higher-order components, such as $F_3$, $F_4$ and $F_5$, have been demonstrated to be much smaller in magnitude comparing to the lower-order components [13]. We have not observed the higher-order components neither.

TABLE II. Effects of $\mathcal{M}'$ and $\mathcal{M}$ on SOT. The first two columns are adapted from Ref. [13].

| $\mathcal{M}$ | | $\mathcal{M}'$ | |
|---|---|---|---|
| $E \perp \mathcal{M}$ | $E \parallel \mathcal{M}$ | $E \perp \mathcal{M}'$ | $E \parallel \mathcal{M}'$ |
| $E \to -E$ | $E \to E$ | $E \to -E$ | $E \to E$ |
| $\tau \to -\tau$ | $\tau \to -\tau$ | $\tau \to \tau$ | $\tau \to \tau$ |
| $\phi_H \to -\phi_H$ | $\phi_H \to \pi - \phi_H$ | $\phi_H \to \pi - \phi_H$ | $\phi_H \to -\phi_H$ |
| $\tau(-\phi_H) = \tau(\phi_H)$ | $\tau(\phi_H) = -\tau(\pi - \phi_H)$ | $\tau(\phi_H) = -\tau(\pi - \phi_H)$ | $\tau(-\phi_H) = \tau(\phi_H)$ |
| $\phi_H$ independent term allowed | $\phi_H$ independent term forbidden | $\phi_H$ independent term forbidden | $\phi_H$ independent term allowed |

B. Absence of $\tau_{AC}$ in control samples without magnetic asymmetry.

In order to verify the effects of magnetic asymmetry, we try to observe $\tau_{AC}$ in two control samples: γ-IrMn$_3$(22)/Py(13) and polycrystalline Pt(10)/Py(10). In γ-IrMn$_3$, the magnetic asymmetry is absent due to disordered magnetic structure, and Pt is known to have negligible magnetic moment let alone any magnetic asymmetry. As expected, $\tau_{AC}$ is absent in both cases for any choice of $\phi_E$. Fig. 10 shows the typical spectra of voltages for the two control samples. Both $V_S$ and $V_A$ shows the usual $\sin(2\phi_H)\cos(\phi_H)$ dependence. This shows that $\tau_{AC}$ is strongly correlated with the magnetic asymmetry.

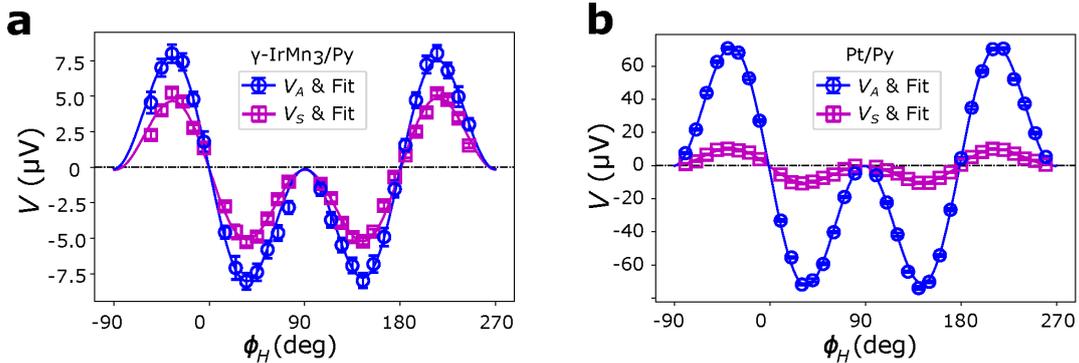

FIG. 10. (a) Absence of the out-of-plane $\tau_{DL}$ in samples without magnetic asymmetry, such as γ-IrMn$_3$ (22)/Py (13) with $\phi_E = 45°$ and (b) Pt (10)/Py (10).

### C. Microscopic origin of the out-of-plane $\tau_{DL}$

In Sec. IIIC, the microscopic origin of $\tau_{AC}$ is correlated with the out-of-plane spin polarization $s_z$, which can be rationalized from both the SHE and the REE perspectives. If $\tau_{AC}$ is generated via the SHE, at least one of $\sigma^z_{zx}$ and $\sigma^z_{zy}$ in the spin Hall conductivity ($\sigma^k_{ij}$) tensor should be non-zero, where $i$, $j$ and $k$ indicate the orientations of spin current, electric current and spin polarization, respectively. We find $\sigma^z_{zx} = 68 \frac{\hbar}{e} \frac{S}{cm}$ for L1$_0$-IrMn [22] and $\sigma^z_{zx} = 95 \frac{\hbar}{e} \frac{S}{cm}$ for L1$_2$-IrMn$_3$ [8], both are more than 50% of their respective $\sigma^y_{zx}$ that accounts for the usual in-plane $\boldsymbol{\tau_{DL}}$.

On the other hand, if $\tau_{AC}$ is generated via the REE, the inversion symmetry in the IrMn/Py bilayer should be broken. Fig. 11 shows that the non-magnetic symmetry of the magnetic sites is *4mm* for both L1$_0$-IrMn and L1$_2$-IrMn$_3$, which indicates that the inversion symmetry is broken along the *c*-axis. In a 2D Rashba model for the *4mm* point group [21,37], the spin polarization $\boldsymbol{s}$ induced by an electric field $\boldsymbol{E}$ can be calculated using the Kubo linear response formalism ($\boldsymbol{s} = \chi \boldsymbol{E}$, where $\chi$ is the response tensor) [21]. When magnetic moment is not considered, $\chi$ only has the usual $x_{21}$ term (meaning an $E \parallel x$ induces an $s \parallel y$), which accounts for the usual $S_A$ and $A_A$ terms in the ST-FMR measurement. It is the contribution from magnetic moments that generates the non-zero $x_{31}$ and $x_{32}$ [21], which are responsible for the $s_z$. Moreover, in the context of REE-based SOT, AFM of different magnetic structures but same symmetry (*4mm*) can generate SOT of similar symmetry and strength [21]. This is consistent with our observation that $\tau_{AC}$ is induced by the common broken $\mathcal{M}'$ in L1$_0$-IrMn and L1$_2$-IrMn$_3$ despite their different magnetic structures. The non-magnetic symmetry of magnetic sites in the 2D Rashba model, though differed from the magnetic symmetry shown in Fig. 1(b), is useful to analyze the effects of magnetic asymmetry on SOT since it demonstrates the symmetry-breaking by magnetic moments.

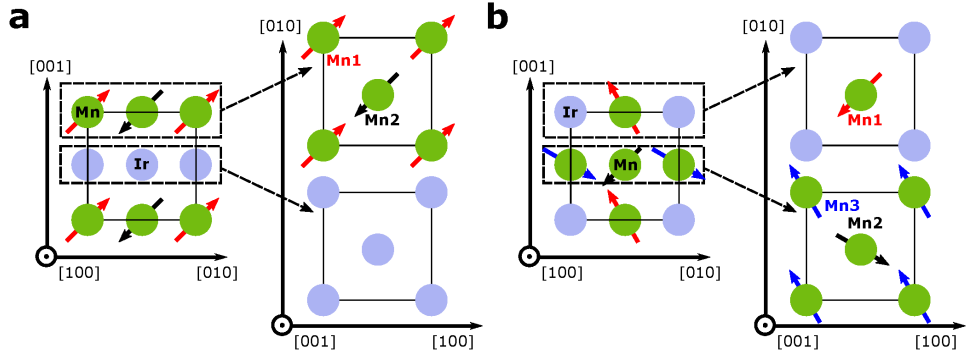

FIG. 11. (a) Interfacial symmetry in L1$_0$-IrMn and (b) L1$_2$-IrMn$_3$.

## V. CONCLUSION

The observed out-of-plane $\tau_{DL}$ is different from previously reported incidences that are only phenomenologically similar. In WTe$_2$, the low crystal symmetry is responsible for a net $s_z$ [13]. However, the crystal symmetries of both L1$_0$-IrMn ($P4/mmm$) and L1$_2$-IrMn$_3$ ($Pm\bar{3}m$) are high and their (001) planes have four-fold rotational symmetry (Fig. 11). In FM systems, the unusual $s_z$ has been attributed to the spin-orbit precession [12], the spin rotation symmetry [14], the AMR effect and the anomalous Hall effect [15,16,38]. These processes require the net magnetization of FM and $s_z$ is very sensitive to interfaces. In contrast, we have verified that IrMn as an AFM has no net magnetization (not shown) and, as shown in Fig. 7(f), inserting a Cu layer between Py and IrMn does not significantly change the strength of $\tau_{AC}$.

In conclusion, we have measured the SOT efficiencies of three phases of IrMn and demonstrated an out-of-plane $\tau_{DL}$ in both L1$_0$-IrMn/Py and L1$_2$-IrMn$_3$/Py. This anomalous SOT is strongly correlated with the magnetic asymmetry in IrMn and is consistent with predictions of the SHE and the REE considering an out-of-plane spin polarization. This out-of-plane spin polarization is favorable for electrical switching of a perpendicularly magnetized ferromagnet, which is compatible with the contemporary high-density magnetic memories. Moreover, unlike FM-based spintronics, AFM materials enjoy many advantages, such as zero stray field, higher stability and faster dynamics [1,21,26]. Therefore, the possibility to control SOT via magnetic asymmetry of AFM not only helps to understand the origin of SOT, but also fuels the development of next-generation AFM spintronics.

## ACKNOWLEDGEMENTS

The Laboratory is a National Research Infrastructure under the National Research Foundation (NRF) Singapore. J. C. is a member of Singapore Spintronics Consortium (SG-SPIN). This research is supported by the Singapore Ministry of Education MOE2018-T2-2-043, AMEIRG18-0022, A*STAR IAF-ICP 11801E0036, and MOE Tier1 R-284-000-195-114. This project was partially supported by the National Key Research and Development Program of China [Title: Nonvolatile and programmable spin logics based on magnetic heterostructures; Grant No.: 2017YFA0206200] for the device micro-fabrications. We would like to acknowledge Singapore Synchrotron Light Source (SSLS) for providing the facility necessary for conducting the research. P. Y. was supported by SSLS via NUS Core Support C-380-003-003-001. A portion of this research used resources at the High Flux Isotope Reactor and Spallation Neutron Source, US-DOE Office of Science User Facilities operated by the Oak Ridge National Laboratory. B.Y. acknowledges the financial support by a research grant from the Benoziyo Endowment Fund for the Advancement of Science. This manuscript has been authored by UT-Battelle, LLC under Contract No. DE-AC05-00OR22725 with the U.S. Department of Energy. The United States Government retains and the publisher, by accepting the article for publication, acknowledges that the United States Government retains a non-exclusive, paid-up, irrevocable, world-wide license to publish or reproduce the published form of this manuscript, or allow others to do so, for United States Government purposes. The Department of Energy will provide public access to these results of federally sponsored research in accordance with the DOE Public Access Plan (http://energy.gov/downloads/doe-public-access-plan).

## REFERENCES

[1] A. Manchon, J. Železný, I. M. Miron, T. Jungwirth, J. Sinova, A. Thiaville, K. Garello, P. Gambardella, *Rev. Mod. Phys.* 91, 035004 (2019).

[2] J. Sinova, S. O. Valenzuela, J. Wunderlich, C. H. Back, T. Jungwirth, *Rev. Mod. Phys.* 87, 1213 (2015).

[3] A. Manchon, S. Zhang, *Phys. Rev. B,* **78**, 212405 (2008).

[4] L. Liu, T. Moriyama, D. C. Ralph, R. A. Buhrman, *Phys. Rev. Lett.* 106, 036601 (2011).

[5] I. M. Miron, G. Gaudin, S. Auffret, B. Rodmacq, A. Schuhl, S. Pizzini, J. Vogel, P. Gambardella, *Nat. Mater.* 9, 230-234 (2010).

[6] Y. Fan, P. Upadhyaya, X. Kou, M. Lang, S. Takei, Z. Wang, J. Tang, L. He, L. Chang, M. Montazeri, G. Yu, W. Jiang, T. Nie, R. N. Schwartz, Y. Tserkovnyak, K. L. Wang, *Nat. Mater.* 13, 699-704 (2014).

[7] C. Pai, L. Liu, Y. Li, H. W. Tseng, D. C. Ralph, R. A. Buhrman, *Appl. Phys. Lett.* 101, 122404 (2012).

[8] W. Zhang, W. Han, S. Yang, Y. Sun, Y. Zhang, B. Yan, S. S. P. Parkin, *Sci. Adv. 2,* e1600759 (2016).

[9] W. Zhang, M. B. Jungfleisch, W. Jiang, J.E. Pearson, A. Hoffmann, F. Freimuth, Y. Mokrousov, *Phys. Rev. Lett.* 113, 196602 (2014).

[10] P. Wadley, B. Howells, J. Železný, C. Andrews, V. Hills, R. P. Campion, V. Novák, K. Olejník, F. Maccherozzi, S. S. Dhesi, S. Y. Martin, T. Wagner, J. Wunderlich, F. Freimuth, Y. Mokrousov, J. Kuneš, J. S. Chauhan, M. J. Grzybowski, A. W. Rushforth, K. W. Edmonds, B. L. Gallagher, T. Jungwirth, *Science*, 351, 6273, 587-590 (2016).

[11] D. Fang, H. Kurebayashi, J. Wunderlich, K. Vyborny, L. P. Zarbo, R. P. Campion, A. Casiraghi, B. L. Gallagher, T. Jungwirth, A. J. Ferguson, *Nat. Nanotechnol.* 6, 413 (2011).

[12] S. C. Baek, V. P. Amin, Y. W. Oh, G. Go, S. J. Lee, G. H. Lee, *Nat. Mater.* 17, 509 (2018).

[13] D. MacNeill, G. M. Stiehl, M. H. D. Guimaraes, R. A. Buhrman, J. Park & D. C. Ralph, *Nat. Phys.* 13, 300-305 (2017).

[14] A. M. Humphries, T. Wang, E. R.J. Edwards, S. R. Allen, J. M. Shaw, H. T. Nembach, J, Q. Xiao, T.J. Silva, X. Fan, *Nat. Comm.* 8, 911 (2017).

[15] A. Bose, D. D. Lam, S. Bhuktare, S. Dutta, H. Singh, Y. Jibiki, M. Goto, S. Miwa, A. A. Tulapurkar, *Phys. Rev. Appl.* 9, 064026 (2018).

[16] Y. Ou, Z. Wang, C. S. Chang, H. P. Nair, H. Paik, N. Reynolds, D. C. RalphDavid, A. MullerDarrell, G. SchlomRobert, A. Buhrman, *Nano Lett.* 19, 6, 3663-3670 (2019).

[17] M. H. D. Guimarães, G. M. Stiehl, D. MacNeill, N. D. Reynolds, D. C. Ralph, *Nano Lett.* 18, 2, 1311-1316 (2018).

[18] J. Zhou, J. Qiao, A. Bourne, W. Zhao, *Phys. Rev. B* 99, 060408 (R) (2019).

[19] G. M. Stiehl, R. Li, V. Gupta, I. E. Baggari, S. Jiang, H. Xie, L. F. Kourkoutis, K. F. Mak, J. Shan, R. A. Buhrman, D. C. Ralph, *Phys. Rev. B* 100, 184402 (2019).

[20] D. E. Laughlin, M. A. Willard, and M. E. McHenry, *Phase Transformations and Evolution in Materials*, The Minerals , Metals and Materials Society, p. 121-137 (2000).

[21] J. Železný, H. Gao, A. Manchon, F. Freimuth, Y. Mokrousov, J. Zemen, J. Mašek, J. Sinova, T. Jungwirth, *Phys. Rev. B* 95, 014403 (2017).

[22] J. Zhou, X. Wang, Y. Liu, J. Yu, H. Fu, L. Liu, S. Chen, J. Deng, W. Lin, X. Shu, H. Y. Yoong, T. Hong, M. Matsuda, P. Yang, S. Adams, B. Yan, X. Han, J. S. Chen, *Sci. Adv. 5,* eaau6696 (2019).

[23] P. Wadley, S. Reimers, M. J. Grzybowski, C. Andrews, M. Wang, J. S. Chauhan, B. L. Gallagher, R. P. Campion, K. W. Edmonds, S. S. Dhesi, F. Maccherozzi, V. Novak, J. Wunderlich, T. Jungwirth , *Nat. Nanotechol,* 13, 362-365 (2018).

[24] O. G. Shpyrko, E. D. Isaacs, J. M. Logan, Yejun Feng, G. Aeppli, R. Jaramillo, H. C. Kim, T. F. Rosenbaum, P. Zschack, M. Sprung, S. Narayanan, A. R. Sandy, *Nature*, 447, 68-71 (2007).

[25] T. Zhao, A. Scholl, F. Zavaliche, K. Lee, M. Barry, A. Doran, M. P. Cruz, Y. H. Chu, C. Ederer, N. A. Spaldin, R. R. Das, D. M. Kim, S. H. Baek, C. B. Eom, R. Ramesh, *Nat. Mater.* 5, 823-829 (2006).

[26] E. V. Gomonay, V. M. Loktev, *Low Temp. Phys.* 40 , 1 (2014).

[27] S. Zhang, P. M. Levy, A. Fert, *Phys. Rev. Lett.* 88, 23 (2002).

[28] M. Harder, Z. X. Cao, Y. S. Gui X. L. Fan, C. M. Hu, *Phys. Rev. B*, 84, 054423 (2011).

[29] S. Fukami, C. Zhang, S. DuttaGupta, A. Kurenkov, H. Ohno, *Nat. Mater.* 15, 535-541 (2016).

[30] A. Kohn, A. Kovacs, R. Fan, G. J. Mclntyre, R. C. C. Ward & J. P. Goff, *Sci. Rep.* 3, 2412 (2013).

[31] I. Tomenoi, H. N. Fuke, H. Iwasaki, M. Sahashi, Y. Tsunoda, *J. Appl. Phys.* 86, 7 (1999).

[32] E. Kren, G. Kadar, L. Pal, J. Solyom, P. Szabo, T. Tarnczi, *Phys. Rev. 171,* 2 : 574 (1968).

[33] W. Zhang, M. B. Jungfleisch, F. Freimuth, W. Jiang, J. Sklenar, J. E. Pearson, J. B. Ketterson, Y. Mokrousov, A. Hoffmann, *Phys. Rev. B* 92, 144405 (2015).

[34] H. Saglam, J. C. Rojas-Sanchez, S. Petit, M. Hehn, W. Zhang, J. E. Pearson, S. Mangin, A. Hoffmann, *Phys. Rev. B* 98, 094407 (2018).

[35] V. Tshitoyan, C. Ciccarelli, A. P. Mihai, M. Ali, A. C. Irvine, T. A. Moore, T. Jungwirth, A. J. Ferguson, *Phys. Rev. B* 92, 214406 (2015).

[36] X. Shu, J. Zhou, J. Deng, W. Lin, J. Yu, L. Liu, C. Zhou, P. Yang, J. Chen, *Phys. Rev. Materials* 3, 114410 (2019).

[37] J. Železný, H. Gao, K. Výborný, J. Zemen, J. Mašek, A. Manchon, J. Wunderlich, *Phys. Rev. Lett.* 113, 157201 (2014).

[38] T. Taniguchi, J. Grollier, M. D. Stiles, *Phys. Rev. Lett*. 3, 044001 (2015).